\documentclass[11pt,twoside]{article}
\usepackage{asp2010}
\resetcounters

\begin{document}

\title{Evidence for IMF Variations from the Integrated Light of SDSS Galaxies}
\author{Erik~A.~Hoversten$^1$, and Karl Glazebrook$^2$
\affil{$^1$Pennsylvania State University, 525 Davey Laboratory, University Park, PA 16802}
\affil{$^2$Centre for Astrophysics \& Supercomputing, Swinburne University, Hawthorn VIC, Australia}}

\begin{abstract}
The H$\alpha$ equivalent width (EW) is the ratio of the H$\alpha$ flux to the continuum at 6565\AA.  In normal star forming galaxies the H$\alpha$ flux is dominated by reprocessed photons from stars with masses greater than 10 M$_\odot$ and the 6565\AA\ continuum is predominantly due to $0.7-3.0$ M$_\odot$ red giant stars.  In these galaxies the H$\alpha$ EW is effectively the ratio of high mass to low mass stars and is thus sensitive to the stellar initial mass function (IMF).  In \citet{HG08} we used $\sim131,000$ galaxies from the Sloan Digital Sky Survey to show evidence for systematic variations in the IMF with galaxy luminosity.  In this proceeding we use that sample, with the addition of H$\delta_A$ measurements, to investigate other parameterizations of the IMF.  We find evidence for IMF variations with surface brightness, and also show that, modulo uncertainties in spectral synthesis models, that 120 M$_\odot$ stars are important in accounting for the observed H$\alpha$ EW distribution.

\end{abstract}

\section{Introduction}

The stellar initial mass function (IMF) has historically been treated as universal across environments and time given that the large errors on the IMF slope from individual cluster measurements are all consistent with a single slope \citep{Kroupa01}.  Recent studies which infer the slope of the IMF from the integrated light properties of galaxies have provided evidence for systematic variations in the IMF.  \citet[][ hereafter HG08]{HG08} found evidence for a link between galaxy luminosity and IMF slope based on a statistical analysis of the $g-r$ colors and H$\alpha$ equivalent widths (EW) of $\sim$ 131,000 galaxies from the Sloan Digital Sky Survey \citep[SDSS;][]{York00}.  They found that the IMF slope was steeper in low luminosity galaxies.  \citet[][ hereafter M09]{M09} found that in a sample of HI selected galaxies the extinction corrected H$\alpha$ to far-ultraviolet (FUV) flux ratio is correlated with galaxy luminosity which they attribute to IMF variations in agreement with HG08.  However, they found that the correlation was stronger with galaxy surface brightness which they argue is the more physically fundamental connection.  \citet{Lee09} compared star formation rates (SFR) calculated from FUV and H$\alpha$ flux for a sample of galaxies spanning five orders of magnitude in SFR.  They found that with decreasing galaxy luminosity the H$\alpha$ SFR systematically underestimated the FUV SFR.  One explanation for their results is a systematically varying IMF.

This proceeding expands on HG08, utilizing the same data set and techniques except where denoted.  For an in depth description the reader is directed to HG08, however a brief overview is provided here.  The HG08 sample contains 130,602 star forming galaxies with H$\beta$ S/N $>5$ selected from SDSS DR4 \citep{DR4}.  However H$\alpha$ measurements, as well as the H$\delta_{\rm A}$, were not from the standard SDSS pipeline, but rather calculated by \citet{CAT}.  Values are extinction corrected and $K$-corrected to $z=0.1$.  A large grid of galaxy spectral synthesis models were calculated over a range of ages, star formation histories (SFH), metallicities, and IMF slopes $\Gamma$ using the publicly available P\'{E}GASE models \citep{PEGASE}.

HG08 compared the $(g-r)_{0.1}$ color to the H$\alpha$ EW, a method pioneered by \citet{K83} and \citet{KTC94}.  In this parameter space IMF variations dominate variations in metallicity and SFH time constant and are largely orthogonal to variations in extinction and age as shown in Figure 3 of HG08.  The H$\alpha$ EW is the ratio of flux in the H$\alpha$ line, which in star forming galaxies is due to reprocessed ionizing photons from young stars with masses greater than 10$M_\odot$, and the continuum at 6565\AA\ which is dominated by red giant stars with masses in the 0.7-3 $M_\odot$ range.  This is effectively the ratio of the number of massive stars to those around a solar mass and is thus sensitive to the IMF.

In HG08 the IMF is parameterized as
\begin{equation}
\frac{{\rm d}n}{{\rm d}\log m} \propto \left \{ \begin{array}{ccc} -0.5 & \mbox{
for}& 0.1 <m/{\rm M}_\odot < 0.5\\
-\Gamma & \mbox{for}& 0.5 < m/{\rm M}_\odot < 120 \end{array} \right.
\label{eq:imfbaldry}
\end{equation}
following \citet{BG03}.  For a Salpeter-like IMF, which is surprisingly consistent with modern values,  $\Gamma=1.35$, however the original Salpeter IMF was a single power-law with no break.

In this proceeding we consider two additional IMF parameterizations.  As discussed in HG08, in this color vs. H$\alpha$ EW parameter space the IMF models themselves can be degenerate.  In Equation \ref{eq:imfbaldry} the IMF slope, $\Gamma$, is allowed to vary, while the upper mass cutoff is held fixed.  Another option, described by
\begin{equation}
\frac{{\rm d}n}{{\rm d}\log m} \propto \left \{ \begin{array}{ccc} -0.5 & \mbox{
for}& 0.1 <m/{\rm M}_\odot < 0.5\\
-1.35 & \mbox{for}& 0.5 < m/{\rm M}_\odot < M_{up} \end{array} \right.
\label{eq:imfmup}
\end{equation}
is to keep the slope at a Salpeter value ($\Gamma=1.35$), but allow the upper mass cutoff, $M_{up}$, to float from 20 to 150 ${\rm M}_\odot$.  For $M_{up} \leq 120\ {\rm M}_\odot$ the models can be explicitly calculated with the P\'{E}GASE models.  Above 120 ${\rm M}_\odot$ the values are extrapolated.

\begin{figure}[!t]
\plotone{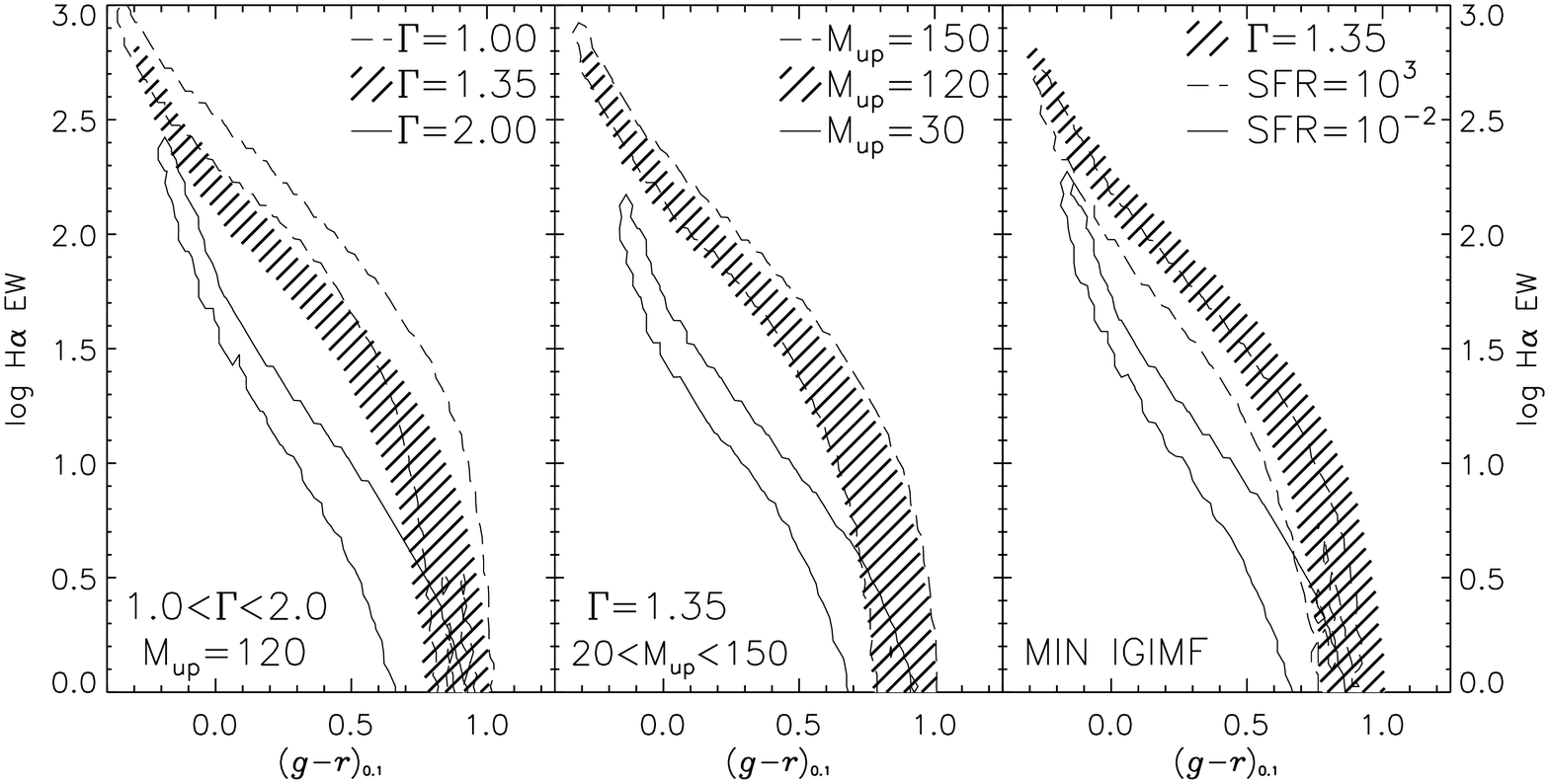}
\label{fig:models}
\caption{Model $(g-r)_{0.1}$ vs. H$\alpha$ EW tracks for several IMF parameterizations.  Each contour represents the area of parameter space covered by the 18,480 models (described in the text) for the given IMF.  In the left panel the slope of the IMF is varied as in Equation \ref{eq:imfbaldry} with an upper mass cutoff of 120 M$_\odot$, in the middle panel $\Gamma=1.35$ while the upper mass cutoff varies following Equation \ref{eq:imfmup}, and in the right panel the minimal IGIMF scenario is used.  The hatched area covers the area of the parameter space for the $\Gamma=1.35$, $M_{up}=120\  {\rm M}_\odot$ models and is identical in all three panels.}
\end{figure}

The second new IMF parameterization is the ``minimal" case of the integrated galaxial initial mass function (IGIMF) theory \citep{WK05}.  In the IGIMF theory the stellar IMF is universal.  However the combined effects of sampling the IMF from many star forming regions, which themselves have a size distribution, leads to an IGIMF which can differ from the universal IMF of stars.  The IGIMF is dependent on the SFR of the galaxy as the maximum stellar cluster mass is related to the galaxy SFR.  The minimal IMF case was calculated for a range of SFRs using the P\'{E}GASE input files of \citet{PAK09}.  A comparison of the three IMF parameterizations can be found in Figure \ref{fig:models}.

The remainder of this proceeding discusses two extensions of the HG08 study.  In Section 2, the addition of H$\delta$ absorption strength to the fitting procedure is discussed.  Section 3 investigates IMF variations with galaxy surface brightness, followed by a discussion of the results in Section 4.

\section{H$\delta_{\rm A}$ Absorption}

Comparing the color to the H$\alpha$ EW is a two parameter fit and thus has inherent limitations.  Changing the age, metallicity, dust extinction, SFH, and IMF can have similar effects.  Fortunately, in the $(g-r)_{0.1}$ vs. H$\alpha$ EW plane used by HG08 the IMF slope largely distinguishes itself from the other effects.  However in Figure 4 of HG08 it can be seen that for galaxies with red colors and low H$\alpha$ EWs the different IMF model tracks become degenerate again.  Of particular importance is the influence of bursts and gasps of star formation which can mimic many other effects. This motivates the use of a third parameter.

Section 7 of HG08 explores the use of the H$\delta$ absorption feature to place further constraints on their results from $(g-r)_{0.1}$ vs. H$\alpha$ EW modeling.  H$\delta$ absorption is due to stellar photospheric absorption lines.  Balmer absorption lines are most pronounced in A stars, and weaken for both hotter and cooler stars following the Saha equation.  For this reason the H$\delta$ absorption can be used as a proxy for the fraction of light being supplied by A stars, and less significantly by B and F stars, in a stellar population.  In a simple stellar population H$\delta$ absorption will peak at around 1 Gyr.  At this time O and B stars have already burned out, but the A stars have yet to leave the main sequence.  As such, H$\delta$ absorption can be used to probe the age of a stellar population as well as uncover evidence of star formation bursts around 1 Gyr in the past.

\citet{Worthey97} set forth two different methods of measuring the H$\delta$ absorption.  As in HG08, we use the H$\delta_{\rm A}$ method which has a wider central bandpass to match the line profiles of A stars.  This is the preferred index for galaxies because it is less noisy in low-S/N spectra and velocity dispersion widens absorptions features.  The H$\delta_{\rm A}$ index ranges from 13 for A4 dwarf stars to -9 in M-type giants.

HG08 found that the distribution of H$\delta_{\rm A}$ was different for their highest and lowest luminosity galaxy bins.  For the most luminous galaxies which are in good agreement with a universal Salpeter IMF the H$\delta_{\rm A}$ values are centered at 3.4 and suggestive of a range of H$\delta_{\rm A}$.  On the other hand the distribution for the faintest luminosity bin is consistent with all galaxies having H$\delta_{\rm A}$=6.1.

When treated separately, as in HG08, H$\delta_{\rm A}$ provides only weak constraints on the IMF.  However, \citet{H08} gave preliminary results demonstrating the power of including H$\delta_{\rm A}$ explicitly in the $(g-r)_{0.1}$ vs. H$\alpha$ EW analysis.  In the three parameter analysis, Equation 8 of HG08 is replaced by
\begin{equation}
\chi_i^2(\Gamma, Z,t,\psi) = \left ( \frac{c_i-c(\Gamma, Z,t,\psi)} 
{\sigma_{c_i}}\right )^2
+ \left ( \frac{w_i-w(\Gamma, Z,t,\psi)}{\sigma_{w_i}}\right )^2
+ \left ( \frac{\delta_i-\delta(\Gamma, Z,t,\psi)}{\sigma_{\delta_i}}\right )^2
\label{eq:chisquare3}
\end{equation}
in the ``pseudo-$\chi^2$" minimization, for a model with IMF $\Gamma$, metallicity $Z$, age $t$, and SFH $\psi$.  For each galaxy $i$, the inputs are the observed $(g-r)_{0.1}$ color $c_i$, H$\alpha$ EW $w_i$, and H$\delta_{\rm A}$ index $\delta_i$ and measurement errors $\sigma_{c_i}$, $\sigma_{w_i}$ and $\sigma_{\delta_i}$.  These are compared to the model values $c(\Gamma,Z,t,\psi)$, $w(\Gamma,Z,t,\psi)$, and $\delta(\Gamma,Z,t,\psi)$.  From these 
the statistical estimator $\chi^2_i(\Gamma,Z,t,\psi)$ value is calculated by brute force for each galaxy, and then marginalized over metallicity, age, and SFH following HG08.

The results of the three-parameter analysis are shown in Figure \ref{fig:mrgbest}.  The top row of Figure \ref{fig:mrgbest} shows the best-fitting IMF models as a function of the $r_{0.1}$ luminosity for each of the three IMF parameterizations.  Each point represents 500 galaxies binned by luminosity, and the lines show the upper and lower 95\% confidence limits on $\Gamma$.  The bottom row shows the goodness of fit parameter, $\overline{\chi^2}$, which is discussed in HG08 and should not be used for any conventional, textbook $\chi^2$ interpretation.

\articlefigure{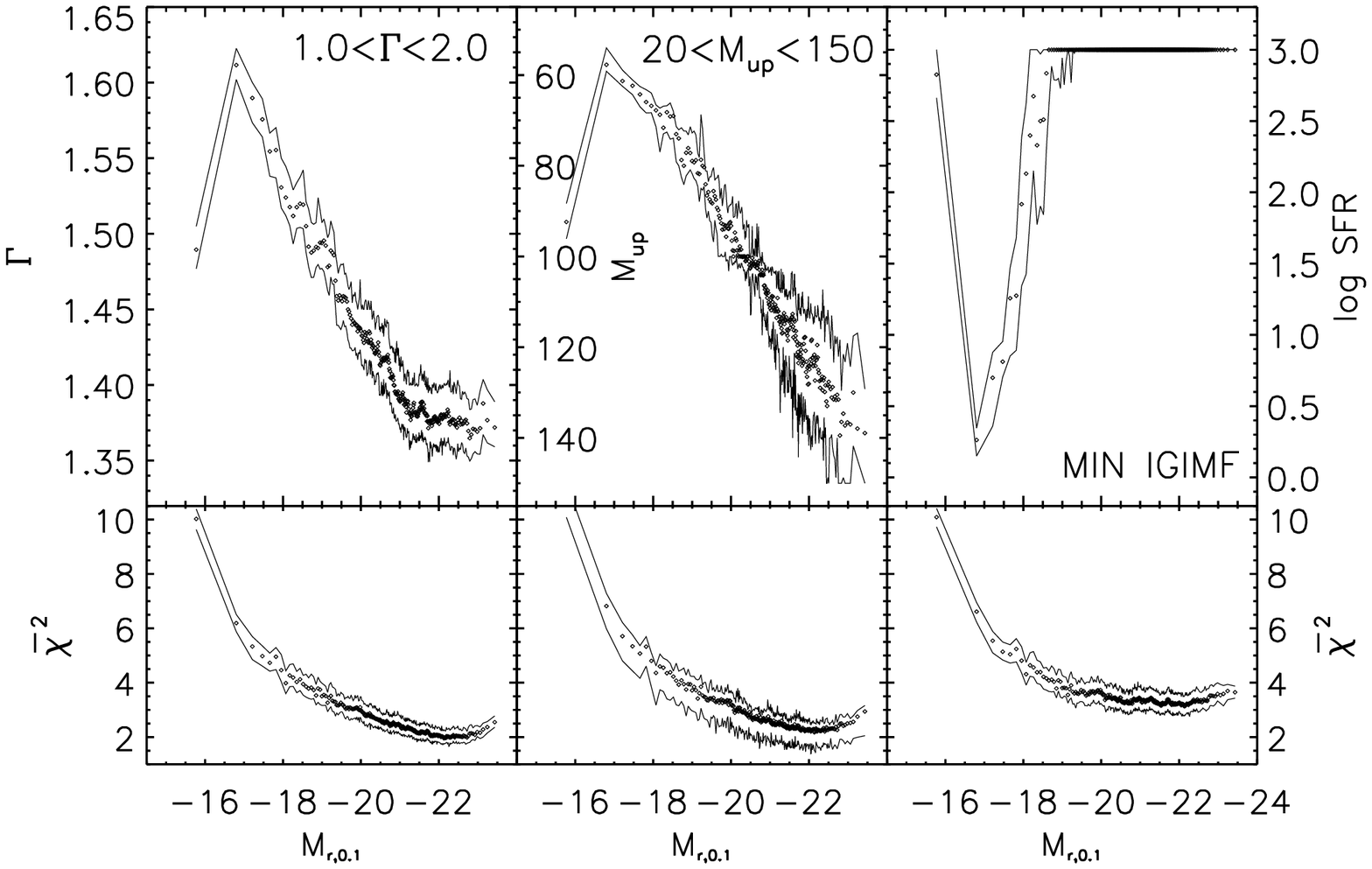}{fig:mrgbest}{MC simulation three parameter results for the full sample binned by ${\rm M}_{r,0.1}$.  As in Figure \ref{fig:models}, the left column gives results for the IMF of Equation \ref{eq:imfbaldry}, at center are the results assuming Equation \ref{eq:imfmup}, and at right are the results for the minimal IGIMF.  Each diamond represents 500 galaxies plotted at the mean ${\rm M}_{r,0.1}$ value of the bin.  The solid lines represent the upper and lower 95\% confidence region measured for each bin.  The three parameter results, including the H$\delta$ absorption data, are plotted in black.  In the top row the best fitting IMF model values as a function of $r$-band luminosity, while in the bottom row the $\overline{\chi^2}$ values for each luminosity bin are plotted.}

Figure \ref{fig:mrgbest} shows that the inclusion of H$\delta_{\rm A}$ has a minimal effect on the derived $\Gamma$ values at low luminosities.  However, on the bright end inclusion of H$\delta_{\rm A}$ has a strong influence on the results.  In the two-parameter fit of HG08, the IMF slope $\Gamma$ steepens at the highest luminosities.  In the three-parameter fit, $\Gamma$ flattens out to nearly the Salpeter slope which much more closely matches what is expected in bright galaxies.  The bottom panel of Figure \ref{fig:mrgbest} shows the quality of fit parameter $\overline{\chi^2}$.  The fit to a single, universal IMF is much better for the high luminosity galaxies, and then deteriorates for the low luminosity galaxies which favor steeper IMF slopes.

Figure \ref{fig:mrgbest} also shows evidence for an increase in $M_{up}$ with increasing galaxy luminosity when using the IMF parameterization of Figure \ref{eq:imfmup}.  For the IGIMF model the SFR increases with galaxy luminosity as expected, however most galaxies are best fit with a SFR of 1000 M$_\odot$ which is a factor of 2 to 3 higher than what is generally accepted for these types of galaxies.

\section{Surface Brightness}

HG08 only investigated IMF variations correlated to galaxy luminosity.  However, M09 has argued that the surface brightness, via the surface mass density, is more fundamentally related to IMF variations than luminosity.  Using the sample and techniques of HG08 it is straightforward to repeat the analysis replacing luminosity with the surface brightness.  This provides an independent check on the surface brightness results in M09 using a completely different technique and galaxy sample.

There are a number of different ways to measure surface brightness in SDSS galaxies.  Here we use the following definition
\begin{equation}
\mu_{r,50} = r - A_r - k_r + 2.5 \log 2 + 2.5 \log \left( \pi r_{50}^2 \right )
\label{eq:surfb}
\end{equation}
where $\mu_{r,50}$ is the $r_{0.1}$ surface brightness within the Petrosian half light radius.  The surface brightness is a function of the SDSS $r$-band Petrosian magnitude $r$, the $r$-band Milky Way extinction $A_r$ measured by \citet{SFD}, the $r$-band $K$-correction to the $z=0.1$ calculated from the code of \citet{Blanton03}, and the $r$-band Petrosian half-light radius $r_{50}$ measured by the SDSS.  

\begin{figure}[!t]
\plotone{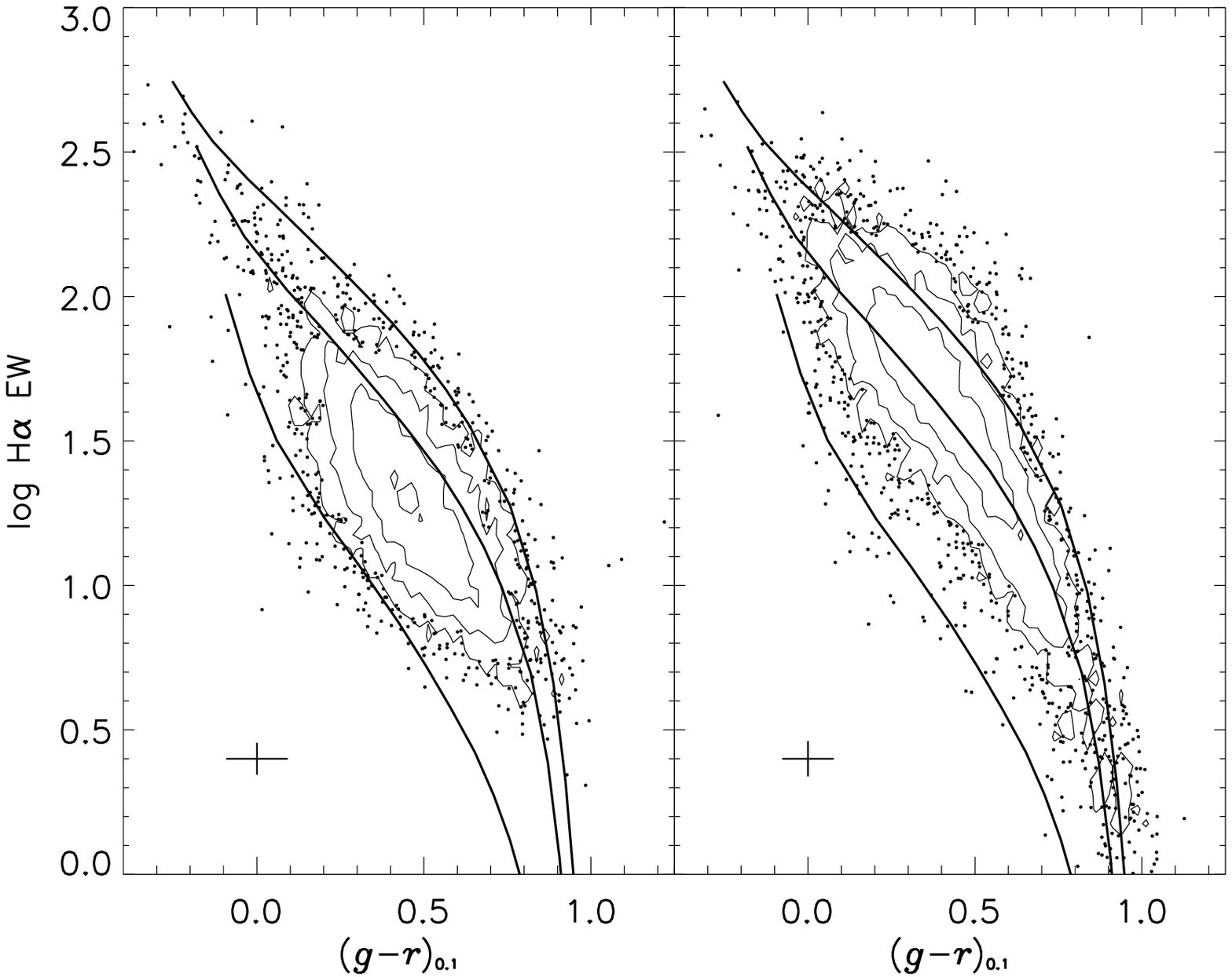}
\label{fig:hagr}
\caption{Distribution in $(g-r)_{0.1}$ - log(H$\alpha$ EW) space for the 10,000 
lowest (left panel) and 10,000 highest (right panel) $\mu_{r,50}$ galaxies in the sample.  Contours are logarithmic.  Outside the last contour individual galaxies are plotted.  Solid lines are model tracks with exponentially decreasing SFHs with $\tau = 1.1$ Gyr and solar metallicity.  The age increases along the tracks from 100 Myr in the upper left to 13 Gyr at the lower right.  The upper line has $\Gamma=1.00$, the middle line is similar to Salpeter's IMF with $\Gamma=1.35$, and the lower line has $\Gamma=2.00$.  The cross in the lower left of each panel indicates the median error bars of the sample.}
\end{figure}

The differences in the H$\alpha$ EWs and colors for the highest and lowest surface brightness galaxies can be seen in Figure \ref{fig:hagr}.  The two panels show the 10,000 lowest (left), and 10,000 highest (right) $\mu_{r,50}$ galaxies in the sample.  The locus of the high surface brightness galaxies tracks the Salpeter IMF model.  However, for fixed $(g-r)_{0.1}$ the locus of the low surface brightness galaxies is clearly lower in H$\alpha$ EW, suggesting a steeper IMF slope.

\begin{figure}[!t]
\plotone{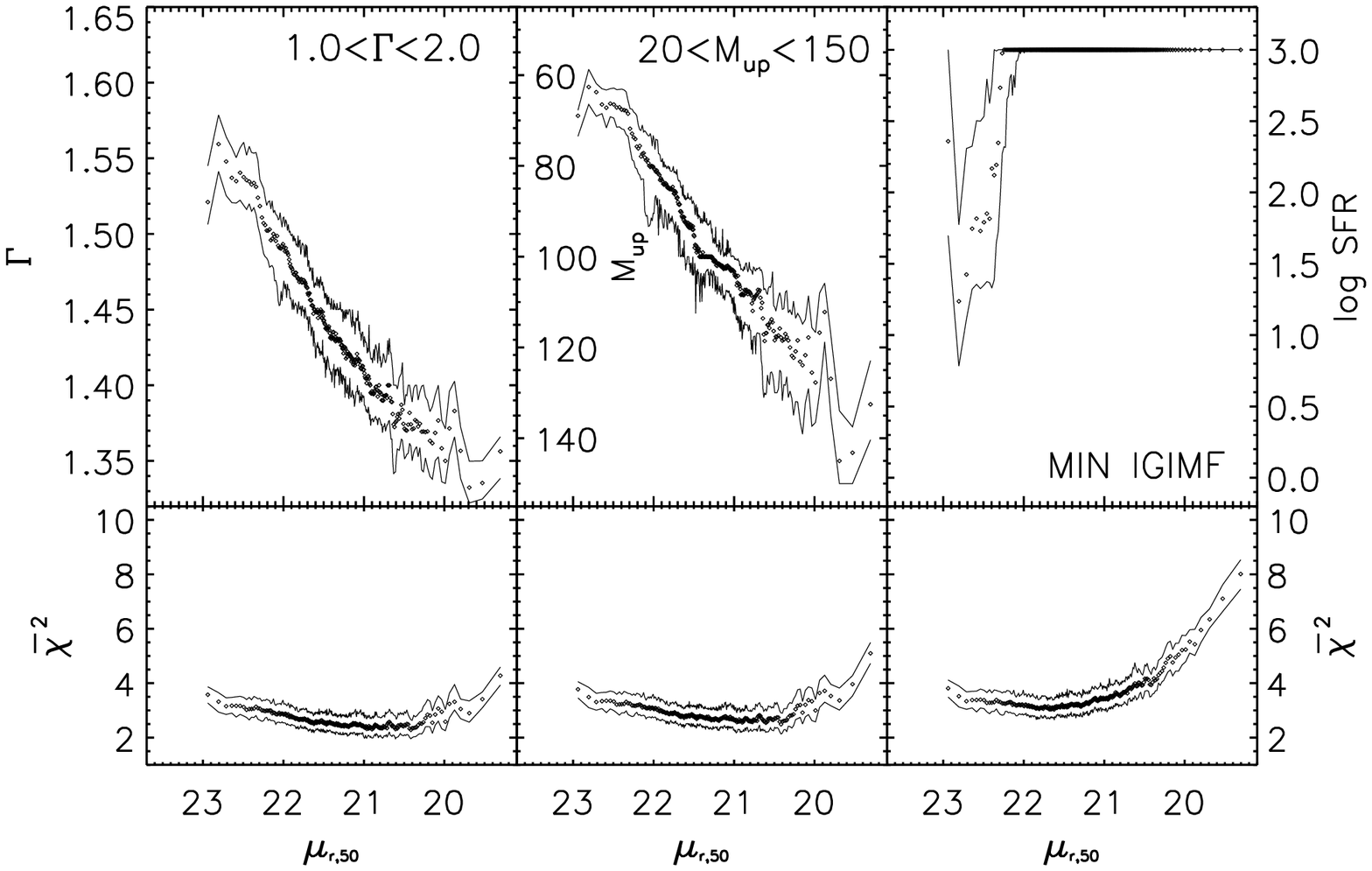}
\label{fig:sbgbest}
\caption{MC simulation three parameter results for the full sample binned by $\mu_{r,50}$. The rest of the description is identical to Figure \ref{fig:mrgbest}.}
\end{figure}

In agreement with the empirical trends of Figure \ref{fig:hagr}, Figure \ref{fig:sbgbest} shows the IMF results as a function of $\mu_{r,50}$ for the three-parameter analysis of Section 2.  The relationship between $\mu_{r,50}$ and $\Gamma$ is nearly linear before flattening out at high surface brightness.  A similar relation is found between surface brightness and $M_{up}$.  As in Figure \ref{fig:mrgbest}, the IGIMF analysis yields SFR values which are implausibly high, although the qualitative trend of increasing SFR with increasing surface brightness is expected.  In a departure from the luminosity based results in Figure \ref{fig:mrgbest} the quality of the fit is much more consistent across the surface brightness bins, particularly for the three-parameter fit.

The SDSS spectroscopic fibers have a fixed size of 3 arcsec.  Given our definition of the surface brightness, bright galaxies with pronounced bulges will have smaller Petrosian half-light radii and will have brighter $\mu_{r,50}$ values.  This is why it is preferable to tailor the aperture to include as much light from the galaxy as possible, as in M09.  The disadvantage is that it is a more labor intensive process; the M09 sample contains 103 galaxies, while ours is over 1,000 time larger which enables more detailed statistical analyses.  Given the nature of this sample disentangling the effects of surface brightness and aperture fraction is a challenge.

In M09 it is argued that surface brightness are more fundamental than the luminosity in IMF variations.  Given our large sample size we can plot the best fitting $\Gamma$ as a function of $r$-band luminosity and surface brightness simultaneously.  This reveals that the variations in $\Gamma$ are not orthogonal to either axis, showing that the IMF variations are a function of both parameters simultaneously.

\section{Discussion}

The analysis presented here shows that in large samples of galaxies there is a clear trend with both galaxy luminosity and surface brightness which cannot be explained using current spectral synthesis models with a universal IMF.  However, there are two primary weaknesses of the results described above.

The most significant limitation of the type of analysis presented here is the treatment of the SFH.  The models described here assume a smoothly varying SFH.  However, as discussed in HG08, SFH discontinuities can have a large impact on the results.  For instance, during a burst of star formation the H$\alpha$ EW will increase and galaxy colors will become bluer which can change the best fitting IMF.  Nonetheless, HG08 demonstrates that while this is an intractable problem for individual galaxies, for statistical samples of galaxies it can be shown that reproducing the observed distribution of galaxies requires an implausible coordination of burst times.  A new analysis of the SFH incorporating the H$\delta_A$ absorption and the new IMF parameterizations is being made for an upcoming paper.

The second problem is that spectral synthesis models have inherent limitations and are still being actively developed.  The systematic effects of the choice of spectral synthesis models is not known for this type of analysis, and the results are subject to change based on improvements in models of stellar evolution.  To quantify this effect, multiple models are being used to determine the scale of variations between models.  The fundamental limitation that spectral synthesis models are only as good as the stellar models that go into them still remains.

However, if variation of the IMF shown here is taken at face value, the results here provide insights into the IMF.  Principle among these is that 120 M$_\odot$ stars are important for creating the observed H$\alpha$ EWs.  Figures \ref{fig:mrgbest} and \ref{fig:sbgbest}  show that the most luminous and highest surface brightness galaxies favor the presence of 120 M$_\odot$ stars when assuming an IMF of the form of Equation \ref{eq:imfmup}.  At the highest luminosities and surface brightnesses the fit is improved by allowing even more massive stars to form.

Secondly, the qualitative behavior of the IGIMF parameterization is what one would expect.  With increasing luminosity and surface brightness the galaxies favor IGIMFs with higher SFRs.  However the calibration of the IGIMF seems to be off as most of the galaxies are best fit by an IGIMF model with a SFR of 1000 M$_\odot$ yr$^{-1}$ which is largely unphysical.  The ``minimal" IGIMF model is truncated at 120 M$_\odot$ and has more massive stars than the standard IGIMF model with 120 M$_\odot$ cutoff \citep{PAK09}.  However even in the minimal case at the highest SFR there are significantly fewer 120 M$_\odot$ stars than in an IMF from Equation \ref{eq:imfbaldry} with $\Gamma=1.35$ or from Equation \ref{eq:imfmup} with $M_{up} = 120 {\rm}_\odot$.  While it needs to be explicitly modeled, it seems that increasing the maximum stellar mass in the IGIMF parameterization beyond $M_{up} = 120 {\rm}_\odot$ could solve this problem.

The results here are also in agreement with M09, in that the IMF appears to vary as a function of galaxy surface brightness.  Lastly, the addition of the h$\delta_A$ data in the analysis improves the constraints on the IMF slope for the most luminous galaxies.

A more in depth analysis of these preliminary results is currently being performed.

\acknowledgements Much of this work was done at Johns Hopkins University with generous support from the Packard Foundation.  EAH acknowledges support from NASA contract NAS5-00136.  KG acknowledges financial support from the Australian Research Council Discovery Project DP 0774469.

\bibliography{hoversten_e}

\end{document}